# SOME ELEMENTARY MECHANISMS FOR CRITICAL TRANSITIONS AND HYSTERESIS IN SIMPLE PREDATOR PREY MODELS

John Vandermeer[i]

**Abstract**

**Trait-mediated indirect effects are increasingly acknowledged as important components in the dynamics of ecological systems. The hamiltonian form of the LV equations is traditionally modified by adding density dependence to the prey variable and functional response to the predator variable. Enriching these non-linear elements with a trait-mediation added to the carrying capacity of the prey creates the dynamics of critical transitions and hysteretic zones.**

The fundamental predator prey models of Lota and Voltera (LV), have long formed the foundation of much of theoretical ecology (Vandermeer and Goldberg, 2013). Other approaches, such as matrix projections (Caswell, 2001 ), interative maps (May, 1980), stochastic processes (Chesson, 1991), reaction-difussion equations (Levin, 1979) have also been important, but the LV approach continues to motivate the development of much theory (Vandermeer, 2004). Independently, the recognition of hysteresis and critical transitions has more recently become an important theme in ecology (Sheffer, 2009), with a major focus on empirical applications (Sheffer et al., 2012). In this note I elaborate some elementary extensions of the basic 2D LV approach in which known ecological factors create background conditions that may lead to hysteresis and critical transitions. Consider the following two dimensional system:

$$\frac{dC}{dt} = B_1(C,R)C - M_1(C,R)C$$

$$\frac{dR}{dt} = B_2(C,R)R - M_2(C,R)R$$

(1)

where C is the predator (consumer) and R is the prey (resource), $B_i$ is the birth rate and $M_i$ is the mortality rate. If we allow $B_1$ to be a simple linear function of R and $M_2$ a simple linear function of C, and $M_1$ and $B_2$ constant, we have:

$$\frac{dC}{dt} = aRC - mC$$

$$\frac{dR}{dt} = bR - aRC$$

(2)

which is to say, the classic quasi-Hamiltonian form of the LV equations. In elementary ecological considerations it is common to begin with the R having a logistic form, when C=0, whence,

---

[i] Department of Ecology and Evolutionary Biology, University of Michigan, Ann Arbor, MI 48109. jvander@umich.edu

$$B_2 - M_2 = r(1 - \frac{R}{K})$$

and thus

$$\frac{dR}{dt} = rR - \frac{r}{K}R^2 \qquad 3$$

It is evident that as R --> 0, d(lnR)/dt --> r, and for any R > 0, d(lnR)/dt < r by a factor of r/K. In ecology the two parameters r and K are referred to as the intrinsic rate of natural increase (herein, the intrinsic rate), and the carrying capacity, respectively. Herein we refer to r/K as the discount rate, for obvious reasons. Formulating $M_2$ as in equation 2, equation 3 expands to,

$$\frac{dR}{dt} = rR - \frac{r}{K}R^2 - aRC = R(r - aRC) - \frac{r}{K}R^2 \qquad 4$$

whence it is evident that this classic form effectively assumes that the predator acting on the system is introduced through the intrinsic rate. There is, of course, no reason why the predator could not operate through the discount rate instead (or in addition to). For example, if the presence of the predator induces the prey to find more places to hide, or stimulates it to seek food at a higher rate, the introduction of the predator into the system can be taken as acting through the discount rate. Ecologically speaking we can formulate this obvious factor by making the carrying capacity an increasing function of the predator, and equation 4 becomes,

$$\frac{dR}{dt} = R(r - aRC) - \frac{r}{K_0 + kC}R^2 \qquad 5$$

Equation 5 represents two effects of the predator, one "direct" effect (parameter a) and one "indirect" effect (parameter k, which we assume is positive -- a negative k does not give interesting results and will not be discussed). The indirect effect here is almost certainly a result of some modification of a "trait" of the prey (e.g. the ability to locate predator-free space) and is thus, in the current terminology popular in ecology, a "trait-mediated indirect effect" (effectively equivalent to the older usage of "higher order effect"). A concrete example is the classical ant/hemipteran mutualism (Zhang et al., 2012) wherein the ant is ultimately a predator of the hemipteran (either directly by consuming, or indirectly by extracting energy from the honeydew secreted by the hemipterans), yet allows the hemipterans to build up very large populations by protecting them from other, more devastating, natural enemies such as parasitic or predatory insects. Thus, this trait-mediated effect changes equation set 2 and we arrive at:

$$\frac{dC}{dt} = aRC - mC$$

$$\frac{dR}{dt} = rR\left(1 - \frac{R}{K_0 + kC}\right) - aRC \qquad 6$$

The zero growth isoclines of system 6 are:

$$R = \frac{m}{a}$$

`                                                                                                    7

$$R = K_0 + \left( k - \frac{a}{r}K_0 \right)C - \frac{ak}{r}C^2$$

whence it is evident that various parameters taken as tuning parameters can produce both critical transition behavior and hysteresis. The phase plane with isoclines and vector field along with a few trajectories is illustrated in figure 1a. Since the predator isocline is R = m/a, using m as the tuning parameter, the visualization is a decreasing intercept (on the R axis) of the predator isocline. Such changes (illustrated in fig 1a) clearly result in a pattern of critical transitions embracing a zone of hysteresis (Fig 1b).

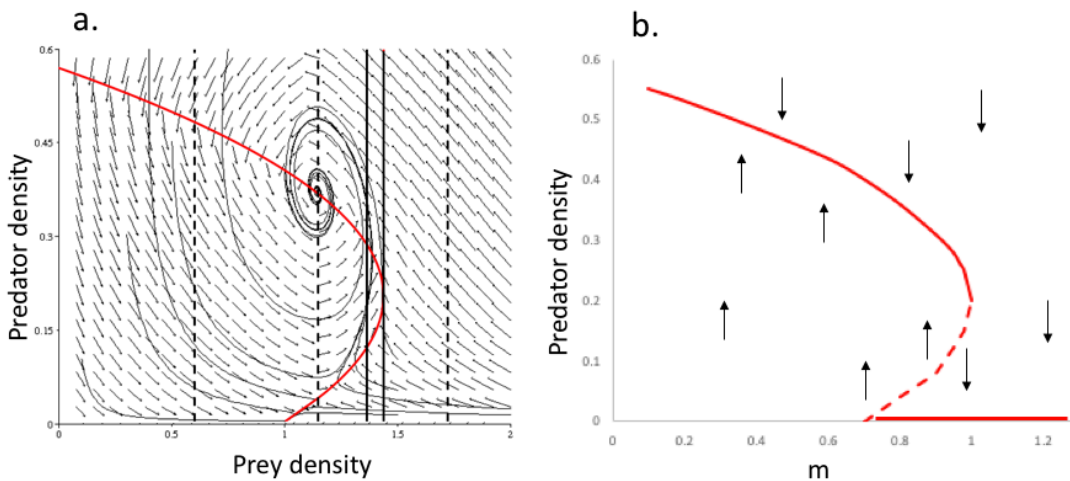

*Figure 1. a. The isoclines, vector field, and illustrative trajectories (equation set 7) of the basic model (equations 6) using m as a tuning parameter. Predator isocline (vertical lines) shown for a = 1.2, 1.003, 0.95, 0.8, and 0.6 respectively from right to left. Clearly the hysteretic zone is between a=1.003 and 0.7, with a saddle/node bifurcation at a=1.003, a stable focus as the upper equilibrium and prey=1 (its carrying capacity) and predator = 0 the lower equilibrium. (Other parameters are, $K_0$ = 1; k = 6, r=0.4). b. Critical transition graph for a.*

Independently of any consideration of critical transitions or hysteresis, Noy Meir (1975) elaborated a graphical approach to the LV model that effectively added three nonlinear components to the basic model. In particular, the prey population was thought to be density dependent (i.e., a carrying capacity was added for the prey), the predator was thought to have a functional response (i.e., at high levels of prey, the predator becomes saturated) and the predator was thought to be dependent on its own density (i.e., some other factor in the environment, other than the prey, limited the predator population at high densities). Following Ong and Vandermeer (2018), with $M_2$ = m, we have,

$$B_2 = r_1(1 - \frac{R}{K_R})$$

and additionally,

$$B_1 = \frac{r_2 R}{1 + bR}(1 - \frac{C}{K_C})$$

$$M_2 = \frac{aC}{1 + bR}$$

whence equation set 1 becomes

$$\frac{dC}{dt} = \frac{r_2 R}{1 + bR}(1 - \frac{C}{K_C})C - mC$$

$$\frac{dR}{dt} = r_1 R(1 - \frac{R}{K_R}) - \frac{aRC}{1 + bR}$$

8

with zero growth isoclines,

$$C = K_C - \frac{mK_C(1 + bR)}{Rr_2}$$

9

$$C = \frac{r_1}{a}\left(1 - \frac{R}{K_R}\right)(1 + bR)$$

illustrated in figure 2.

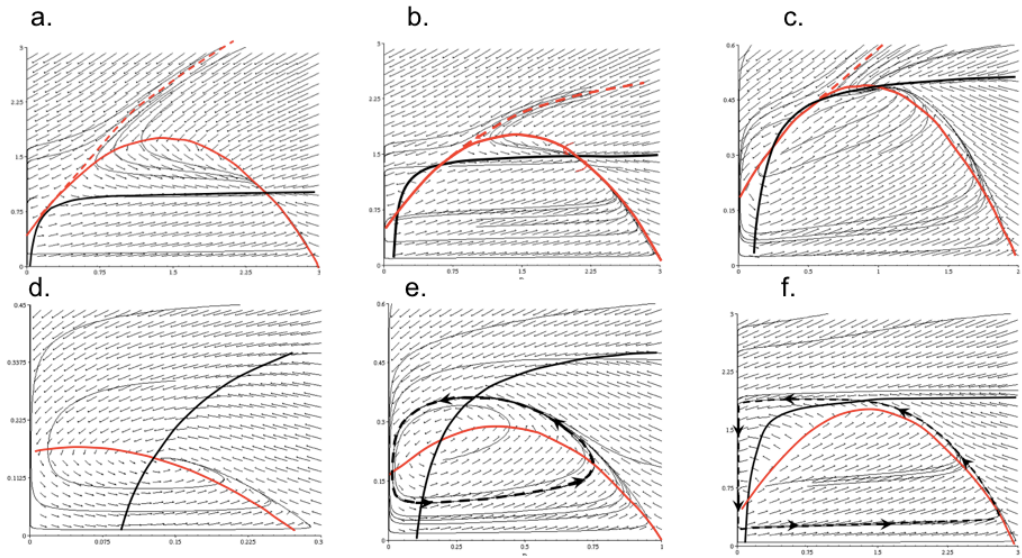

*Figure 2. Exemplary graphic displays of equations 9 (in phase space with zero growth isoclines and vector fields, for various parameter values). In a, b and c, the dashed red curve indicates an unstable inset to the saddle point. In e and f the dashed black curve indicates stable limit cycle. a. parameters $r_2 = 2$; $b = 5$; $K_C = 1.4$; $m = 0.1$; $r_1 = 3.3$; $K_R = 3$; $a = 8$. b. parameters same as in a except $KC = 2$. c. parameters same as in a except $KC = 0.8$; $m = 0.2$; $r1 = 1.28$; $KR = 2$; d. parameters same as in a except $r2 = 3$; $KC = 0.8$; $m = 0.2$; $r1 = 1.28$; $KR = 0.3$; e. parameters same as in a except $r2 = 3$; $KC = 0.8$; $m=0.2$; $r1 = 1.28$; $KR = 1$. f. Stable limit cycle repeatedly approaching $R = 0$, effectively driving R extinct, which is followed by C going extinct also. Parameters same as in a except $KC = 2.72$.*

If the conditions giving rise to equations 8 are expanded to include the trait mediated conditions of equations 6, the following equations arise:

$$\frac{dC}{dt} = \frac{r_2 R}{1 + bR}(1 - \frac{C}{K_C})C - mC$$

$$\frac{dR}{dt} = r_1 R(1 - \frac{R}{(K_R + kC)}) - \frac{aRC}{1 + bR}$$

10

with zero growth isoclines,

$$C = K_C - \frac{mK_C(1 + bR)}{Rr_2}$$

11

$$C(K_R - \frac{r_1 k}{a}) + kC^2 = \frac{r_1}{a}K_R + \left(\frac{r_1 b K_R}{a} - \frac{r_1}{a}\right)R - \frac{r_1 b}{a}R^2 + \frac{r_1 kb}{a}RC$$

The resulting qualitative dynamics are illustrated in figure 3. There are obviously two distinct regions of hysteresis (fig 3a and c), qualitatively corresponding to the previous hysteretic patterns (equations 6 and 9), not surprisingly since equations 10 are a combination of the two mechanisms. Note that the sketch in figure 3d is approximate since it is a consensus of bifurcation diagrams with various values of $C_0$ (the initiation values of C). Properly it should be displayed in 3D, but the basic qualitative nature of the diagram does not change, and it is evident from the dashed red lines (representing the separatrices of the two basins) in figs 3a and c, that the initial value of C will make a difference in the details, but the basic structure of the critical transition graph in figure 3d, will not change much. It is worth noting that the hysteretic patterns suggested in figure 3 occur at distinct points along the range of predator mortality and carrying capacity values (m), similar to that reported elsewhere (Ong and Vandermeer, 2018). With more complicated tuning parameters such hysteretic zones can overlap.

In summary, the basic idea of a trait-mediated indirect interaction can be incorporated as a "higher order interaction" in a variety of ways. Here we explore the consequences of a framework in which the parameter corresponding to the carrying capacity of the prey species is augmented by the presence of a predator. This framing adds a nonlinear component as a divisor of one of the terms, resulting in some complicated oscillatory behavior in the basic model. The existence of critical transitions bordering hysteritic zones emerges with both trait-mediated additions and a classic functional response to the predator. Incorporating functional response along with trait mediation interestingly may result in a dual hysteretic zone, paralleling the zones that emerge from each of those nonlinearities added separately.

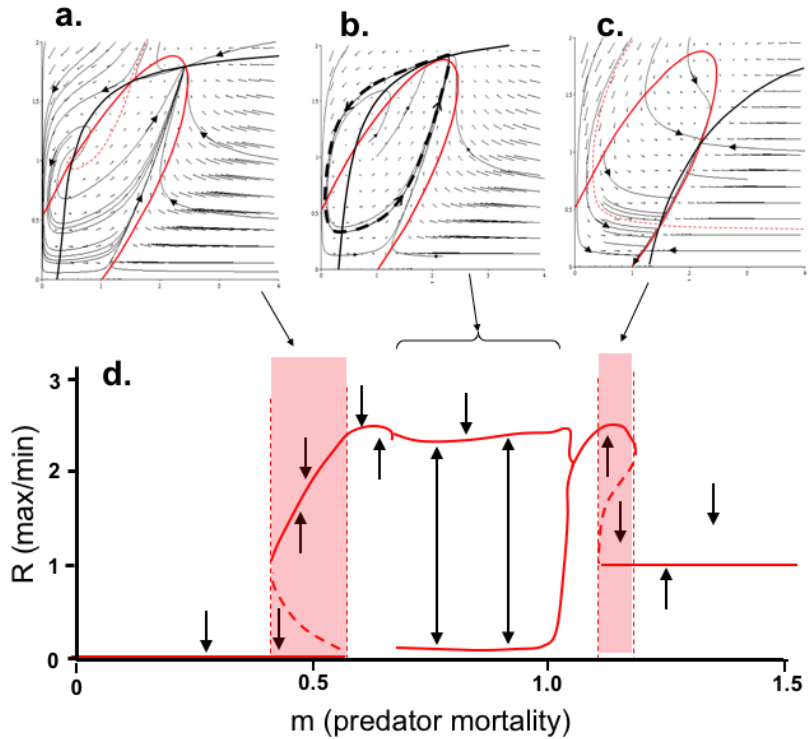

*Figure 3. a,b, and c: Exemplary graphic displays of equations 11 (in phase space with zero growth isoclines and vector fields, for various values of the parameter m). d. Sketch of maximum and minimum resource values at different values of m, where the carrying capacity of the predator is given as 15.8m - 6.73, assuming that there is a tradeoff between mortality and carrying capacity.*

It is worth noting that the hysteretic patterns suggested in figure 3 occur at distinct points along the range of predator mortality and carrying capacity values (m), similar to that reported elsewhere (Ong and Vandermeer, 2018). With more complicated tuning parameters such hysteretic zones can overlap.

# References

Caswell, H. (2001). *Matrix population models*. John Wiley & Sons, Ltd.

Chesson, P. (1991). Stochastic population models. In *Ecological heterogeneity* (pp. 123-143). Springer, New York, NY.

Levin, S. A. (1979). Non-uniform stable solutions to reaction-diffusion equations: applications to ecological pattern formation. In *Pattern formation by dynamic systems and pattern recognition* (pp. 210-222). Springer, Berlin, Heidelberg.

May, R. M. (1980). Nonlinear phenomena in ecology and epidemiology. *Annals of the New York Academy of Sciences*, *357*(1), 267-281.

Noy-Meir, I. (1975). Stability of grazing systems: an application of predator-prey graphs. *The Journal of Ecology*, 459-481.

Ong, T., and Vandermeer, J. 2018. Multiple Hysteretic Patterns from Elementary Population Models. Theoretical Ecology, in press.

Scheffer, M. (2009). *Critical transitions in nature and society*. Princeton University Press.

Scheffer, M., Carpenter, S.R., Lenton, T.M., Bascompte, J., Brock, W., Dakos, V., Van de Koppel, J., Van de Leemput, I.A., Levin, S.A., Van Nes, E.H. , Pascual, M., and Vandermeer, J. (2012). Anticipating critical transitions. *science*, *338*(6105), pp.344-348.

Vandermeer, J. (2004). Coupled oscillations in food webs: balancing competition and mutualism in simple ecological models. *The American Naturalist*, *163*(6), 857-867.

Vandermeer, J. H., & Goldberg, D. E. (2013). *Population ecology: first principles*. Princeton University Press.

Zhang, S., Zhang, Y., & Ma, K. (2012). The ecological effects of the ant–hemipteran mutualism: a meta-analysis. *Basic and Applied Ecology*, *13*(2), 116-124.